\documentclass{aip-cp}

\usepackage{bm}
\usepackage[numbers]{natbib}
\usepackage{rotating}
\usepackage[utf8]{inputenc}

\begin{document}

\title{Jarzynski Equality In The Context Of Maximum Path Entropy}

\author[aff1,aff2]{Diego Gonz\'alez\corref{cor1}}
\author[aff1]{Sergio Davis}

\affil[aff1]{Comisión Chilena de Energía Nuclear, Casilla 188-D, Santiago, Chile.}
\affil[aff2]{Departamento de F\'{\i}sica, Facultad de Ciencia, Universidad de Chile}
\corresp[cor1]{Corresponding author: dgonzalez@gnm.cl}

\maketitle

\begin{abstract}
In the global framework of finding an axiomatic derivation of nonequilibrium Statistical
Mechanics from fundamental principles, such as the maximum path entropy -- also known as Maximum
Caliber principle -- , this work proposes an alternative derivation of the well-known
Jarzynski equality, a nonequilibrium identity of great importance today due to its applications 
to irreversible processes: biological systems (protein folding), mechanical systems, among others. 
This equality relates the free energy differences between two equilibrium thermodynamic states with 
the work performed when going between those states, through an average over a path ensemble.

In this work the analysis of Jarzynski's equality will be performed using the formalism of inference 
over path space. This derivation highlights the wide generality of Jarzynski's original result, which 
could even be used in non-thermodynamical settings such as social systems, financial and ecological systems.
\end{abstract}

\newcommand*{\mycommand}[1]{\texttt{\emph{#1}}}

\section{INTRODUCTION}

Following Jaynes' idea about extending the Maximum Entropy principle~\cite{Jaynes1957} to time-dependent systems via the Maximum Caliber Principle~\cite{Jaynes1980,Presse2013,Gonzalez2016}, it is possible to recover a large part of the structure of 
non-equilibrium statistical mechanics. In particular we can recover all the PDEs that are particular cases of the 
continuity equation~\cite{GonzalezContinuity}, such as the Liouville equation~\cite{GonzalezLiouville}, the Fokker-Planck 
equation, among others. The resulting structure consists of general nonequilibrium relations known as fluctuation theorems. 

In the past two decades several non-equilibrium relations have appeared that describe some 
properties of processes connecting two equilibrium states. Among them, arguably the most renowned 
is the Jarzynski equality (JE)~\cite{Jarzynski1997}. This equality goes beyond the statement of the 
second law of thermodynamics as 

\begin{equation}
W \geq \Delta F,
\end{equation}
where $\Delta F$ is the difference of Helmholtz 
free energy between equilibrium states and $W$ is the macroscopic work needed to move the system from 
one state to the other. A fundamental point is that the JE connects nonequilibrium expectations with 
equilibrium properties. In particular, it asserts that

\begin{equation}
\Big<\exp(-\beta W[\Gamma()])\Big> = \exp(-\beta\Delta F),
\end{equation}
where $W[\Gamma()]$ is the microscopic work needed to transit from a particular microstate in $A$ to 
the corresponding microstate in $B$ along the path $\Gamma(t)$. This equality is remarkable in that 
it has been extensively verified in experiments on individual proteins~\cite{Hummer2001, Liphardt2002, 
Bustamante2005} by performing repeated cycles of stretching at different rates in order to obtain the 
free energy difference between the folded and unfolded thermodynamic states.

\section{SETTING OF THE PROBLEM}

Let us consider a system in a thermodynamic equilibrium state $A(E,V,T,\ldots)$ which is described by the Hamiltonian $H_{A}(\Gamma_{A})$, 
where $\Gamma_{A}$ is a microstate of the system $A$. From this point the system evolves \emph{without any heat exchange}
into an new thermodynamical equilibrium state $B(E',V',T,\ldots)$ described by the Hamiltonian $H_{B}(\Gamma_{B})$, at the
same temperature $T$ (but of course with different thermodynamical properties and different micro-states). Jarzynski's
equality (JE) asserts that the difference in Hemholtz free energy between the systems, $\Delta F= F_{B}-F_{A}$, is related to 
the expectation over paths of an exponential function of the work required to go from the micro-state $\Gamma_{A}$ to
$\Gamma_{B}$, by the equality 

\begin{equation}
\Big<\exp(-\beta W[\Gamma()]) \Big>_I = \exp(-\beta \Delta F).
\label{jarzynski}
\end{equation}

\section{PROOF OF JARZYNSKI EQUALITY}

In order to calculate the expectation over an \emph{ensemble} of paths of the quantity $\exp(-\beta W[\Gamma()])$, first we
need to clarify the definition of that expectation on such a space. Our quantity is given by

\begin{equation}
\Big<\exp(-\beta W[\Gamma()])\Big>_I = \int {D_{\Gamma}} \;\; P[\Gamma()|I] \exp(-\beta W[\Gamma()]) 
\end{equation}
where $P[\Gamma()|I]$ is the probability of the system going through a path $\Gamma()$ given certain information $I$. Now we
wish to include the assumptions that systems $A$ and $B$ are in thermodynamical equilibrium, into the probability distribution 
$P[\Gamma()|I]$. To do this we use the marginalization rule of probability~\cite{Sivia2006} as follows. First,

\begin{equation}
P[\Gamma()|I] = \int_{\Gamma_{A}} \int_{\Gamma_{B}} {d\Gamma_{A}}\;{d\Gamma_{B}}\;\; P(\Gamma(),\Gamma_{A},\Gamma_{B}|I),
\end{equation}
then separating the probability according to the product rule, we obtain

\begin{equation}
P[\Gamma()|I] = \int_{\Gamma_{A}} \int_{\Gamma_{B}} {d\Gamma_{A}}\;{d\Gamma_{B}}\;\; P(\Gamma_{A},\Gamma_{B}|I) P[\Gamma()|\Gamma_{A},\Gamma_{B},I].
\end{equation}
It follows that the path expectation is of the form

\begin{eqnarray}
\Big< \exp(-\beta W[\Gamma()]) \Big>_{I} = \int_{\Gamma_{A}} \int_{\Gamma_{B}} \int {D_{\Gamma}} \;
{d\Gamma_{A}}\;{d\Gamma_{B}}\;\; P(\Gamma_{A},\Gamma_{B}|I) P[\Gamma()|\Gamma_{A},\Gamma_{B},I] \exp(-\beta W[\Gamma()])
\nonumber \\ 
= \Big<\Big< \exp(-\beta W[\Gamma()]) \Big>_{\Gamma_{A},\Gamma_{B},I} \Big>_{I}.
\label{exp_rule}
\end{eqnarray}
If the system evolves from $A$ to $B$ without heat exchange (isolated system), then the work according to the First Law of
Thermodynamics is $W = \Delta U - Q$. Here $W$ is the macroscopic work, defined as the expectation $W=\big<W[\Gamma()]\big>$ of 
the microscopic work $W[\Gamma()] = H_{B}(\Gamma_{B}) - H_{A}(\Gamma_{A})$ with $U_{A} = \Big< H_{A}(\Gamma_{A}) \Big>_{I}$, 
$U_{B} = \Big< H_{B}(\Gamma_{B}) \Big>_{I}$. 

\noindent
We can write Equation \ref{exp_rule} as

\begin{equation}
\Big< \exp(-\beta W[\Gamma()]) \Big>_{I} = \Big< \Big< \exp(-\beta [H_{B}(\Gamma_{B}) - H_{A}(\Gamma_{A})] \Big>_{\Gamma_{A},\Gamma_{B},I} \Big>_{I},
\end{equation}
where the internal expectation is independent of the boundary conditions $\Gamma_{A}$ and $\Gamma_{B}$, and therefore 
can be written as

\begin{equation}
\Big< \exp(-\beta W[\Gamma()]) \Big>_{I} = \Big< \exp(-\beta [H_{B}(\Gamma_{B}) - H_{A}(\Gamma_{A})]) \Big>_{I}.
\end{equation}

\noindent
This means the right-hand side is independent of the path \emph{ensemble} $P[\Gamma()|\Gamma_{A},\Gamma_{B},I]$, it only
depends on the joint probability of the boundary conditions $P(\Gamma_{A},\Gamma_{B}|I)$. Using the product rule of
probability to separate it into two factors we get

\begin{equation}
\Big< \exp(-\beta W[\Gamma()]) \Big>_{I} = \int_{\Gamma_{A}} \int_{\Gamma_{B}} \; {d\Gamma_{A}}\;{d\Gamma_{B}}\;\; 
P(\Gamma_{A}|I) P(\Gamma_{B}|\Gamma_{A},I) \exp(-\beta H_{B}(\Gamma_{B})) \exp(\beta H_{A}(\Gamma_{A})).
\end{equation}

\noindent
If the system starts in an equilibrium thermodynamical state $A$, described by $H_{A}(\Gamma_{A})$, then the probability distribution
for the microstates $\Gamma_{A}$ is, according to the maximum entropy principle, the canonical ensemble 

\begin{equation}
P(\Gamma_{A}|I) = \frac{\exp(-\beta H_{A}(\Gamma_{A}))}{Z_{A}(\beta)},
\end{equation}
where the factor $\exp(-\beta H_{A}(\Gamma_{A}))$ cancels out the term $\exp(\beta H_{A}(\Gamma_{A}))$ in the expectand. 
Finally, we get

\begin{equation}
\Big< \exp(-\beta W[\Gamma()]) \Big>_{I} = \frac{1}{Z_{A}(\beta)} \int_{\Gamma_{A}} \int_{\Gamma_{B}} \; 
{d\Gamma_{A}}\;{d\Gamma_{B}}\;\; P(\Gamma_{B}|\Gamma_{A},I) \exp(-\beta H_{B}(\Gamma_{B})),
\end{equation}
which we can rewrite to introduce the canonical distribution at $B$,

\begin{equation}
P(\Gamma_{B}|I) = \frac{\exp(-\beta H_{B}(\Gamma_{B}))}{Z_{B}(\beta)},
\end{equation}
by multiplying by $Z_{B}(\beta)/Z_{A}(\beta)$. This is because system $B$ can also be described using MaxEnt, due 
to the assumption of thermodynamical equilibrium. Now the right-hand side is of the form 

\begin{eqnarray}
\Big< \exp(-\beta W[\Gamma()]) \Big>_{I} = \frac{Z_{B}(\beta)}{Z_{A}(\beta)} \int_{\Gamma_{A}} \int_{\Gamma_{B}} \; 
{d\Gamma_{A}}\;{d\Gamma_{B}}\;\; P(\Gamma_{B}|\Gamma_{A},I) P(\Gamma_{B}|I) \nonumber \\
= \frac{Z_{B}(\beta)}{Z_{A}(\beta)} \Big< \int_{\Gamma_{A}} \; {d\Gamma_{A}}\;\; P(\Gamma_{B}|\Gamma_{A},I) \Big>_{I}.
\end{eqnarray}

\noindent
In this expression the factor $P(\Gamma_{B}|\Gamma_{A},I)$, the probability of the microstate $\Gamma_{B}$ given the 
initial micro-state was $\Gamma_{A}$ is a Dirac delta for systems following a deterministic dynamics. That is,

\begin{equation}
P(\Gamma_{B}|\Gamma_{A},I) = \delta(\Gamma_{B} - \Gamma),
\end{equation}
where $\Gamma$ is the final microstate of the Hamiltonian dynamics with initial condition $\Gamma_{A}$. According to this, 
$\Gamma$ can be written as

\begin{equation}
\Gamma = \Gamma_{A} + \int_{A}^{B} \;{dt} \;\; \dot \Gamma(\Gamma(t)).
\end{equation}

\noindent
Replacing this in the right-hand side of the expression above, we have

\begin{equation}
\Big< \exp(-\beta W[\Gamma()]) \Big>_{I} = \frac{Z_{B}(\beta)}{Z_{A}(\beta)} \Big< \int_{\Gamma_{A}} \; 
{d\Gamma_{A}}\;\; \delta(\Gamma_{B} - \Gamma_{A} - \int_{A}^{B} \;{dt} \;\; \dot \Gamma) \Big>_{I},
\end{equation}
and therefore

\begin{equation}
\Big< \exp(-\beta W[\Gamma()]) \Big>_{I} = \frac{Z_{B}(\beta)}{Z_{A}(\beta)} = \exp(-\beta\Delta F),
\end{equation}
because, according to the definition of free energy\cite{Callen1985}, $F=-\beta^{-1} \ln Z$. This is 
the Jarzynski equality, and when the Jensen inequality, $\big<\exp(g)\big> \leq \exp(\left<g\right>)$ for any $g$, 
is applied to it, we recover the well-known inequality for macroscopic work,

\begin{equation}
\Delta F \leq \Big< W[\Gamma] \Big>_{I}.
\end{equation}

\section{JARZYNSKI EQUALITY AND MODEL COMPARISON}

As we have seen, the JE is just a consequence of two ``equilibrium states'' (MaxEnt distributions) 
connected via a nonequilibrium process. In principle it can be used to compute the Bayes factor~\cite{MacKay2003}

\begin{equation}
\frac{P(M_A|D, I)}{P(M_B|D, I)}=\frac{P(D|M_A, I)}{P(D|M_B, I)}\frac{P(M_B|I)}{P(M_A|I)}
\end{equation}
if the prior ratio $P(M_B|I)/P(M_A|I)$ is assumed known and we take

\begin{eqnarray}
P(D|M_A, I) = \int d\theta \exp(\ln P(\theta|M_A, I)+\ln P(D|\theta, M_A, I)) = Z_A, \nonumber \\
P(D|M_B, I) = \int d\phi \exp(\ln P(\phi|M_B, I)+\ln P(D|\phi, M_A, I)) = Z_B,
\end{eqnarray}
as partition functions of equilibrium (MaxEnt) problems.

\section{Conclusions}

Because the JE is just a consequence of two ``equilibrium states'' that must be described as MaxEnt distributions, 
it is possible to extend the validity of the result to the Bayesian inference framework almost unmodified, being 
plausible to apply the equality in different subjects of Science such as signal analysis, biology, economy, among others. In
fact, as recently shown by the authors~\cite{Davis2016}, arbitrary inference problems with binary choices are described by the formalism of 
thermodynamical first-order phase transitions and have associated with them concepts such as free energy differences $\Delta
F$.

It is important to note that the validity of this result is independent of the intermediate process, the only important
detail being that evolution from $A$ to $B$ (both equilibrium states) is at constant temperature.

Understanding these and other non-equilibrium relationships from the point of view of Bayesian inference, is a key goal to
aim for in the larger scheme of formalizing a theory for non-equilibrium Statistical Mechanics, being the JE a first step 
into this territory. It shows that the formalism for inference over paths, i.e., the use of the Bayesian probability framework in 
path space, and the maximum entropy principle (including the maximum path entropy principle) seems to be sufficient for the
completion of this objective.  

\section{ACKNOWLEDGMENTS}

DG acknowledges support from CONICYT PhD fellowship 21140914. SD acknowledges support from FONDECYT grant 1140514.

\bibliographystyle{aipnum-cp}
\bibliography{jarzynski}

\end{document}